\documentclass[]{aa}
\usepackage{txfonts}
\usepackage{graphicx}
\usepackage{natbib}
\bibpunct{(}{)}{;}{a}{}{,}    
\usepackage[dvips]{color}
\newcommand{\beq}{\begin{equation}}
\newcommand{\eeq}{\end{equation}}
\newcommand{\bea}{\begin{eqnarray}}
\newcommand{\eea}{\end{eqnarray}}
\newcommand{\pdoverd}[2]{\frac{\partial #1}{\partial #2}}
\newcommand{\pdoverdt}[1]{\frac{\partial #1}{\partial t}}

\newcommand{\Sect}[1]{Sect.~\ref{#1}}

\def\MSOL{{\rm M_\odot}}
\def\RSOL{{\rm R_\odot}}

\def\AU{{\rm AU}}
\def\PC{{\rm pc}}
\begin{document}
\title{Evolution of irradiated Circumbinary Disks}
\author{Richard G\"unther, Christoph Sch\"afer \& Wilhelm Kley}
\offprints{R. G\"unther,\\ \email{rguenth@tat.physik.uni-tuebingen.de}}
\institute{Institut f\"ur Astronomie \& Astrophysik, 
     Abt. Computational Physics,
           Auf der Morgenstelle 10, D-72076 T\"ubingen, Germany}
\date{Received February 10, 2004; accepted April 23, 2004}
\abstract{%
We study the evolution and emission of circumbinary disks around close
classical T Tauri binary systems.  High resolution numerical
hydrodynamical simulations are employed to model a system consisting of a central
eccentric binary star within an irradiated accretion disk.  A detailed
energy balance including viscous heating, radiative cooling and
irradiation from the central star is applied to calculate accurately the
emitted spectral energy distribution.

Numerical simulations using two different methods, the previously developed
Dual-Grid technique with a finite difference discretization, and
the Smoothed Particle Hydrodynamics method are employed to compare the
hydrodynamical features and strengthen our conclusions.

Physical parameters of the setup are chosen to model
the close systems of \object{DQ Tau} and \object{AK Sco}.
Using the self-consistent models, we are able to fit the observed spectral
energy distributions by constraining parameters such as disk mass, density profile
and radial extension for those systems.
We find that the incorporation of irradiation effects is necessary to
obtain correct disk temperatures.
}
\maketitle
\section{Introduction}
From observations it is known that a large fraction of young
stars are in binaries. The formation of binary systems by close
interactions in a star forming region was studied numerically by
\citet{2002MNRAS.336..705B}.
Typically, a circumbinary accretion disk surrounding the double star is formed
during this process.
Here we concentrate on a later phase and model the hydrodynamic
evolution of the circumbinary disk around close binary stars having a
separation of only a fraction of an AU.  This is inspired by
observations of the spectroscopic binaries DQ Tau and AK Sco for which
there is also a spectral energy distribution (SED) available. The
latter system AK Sco was reanalyzed recently by
\citet{2003A&A...409.1037A}.

In a previous paper \citep{2002A&A...387..550G} we
modeled the evolution of circumbinary disks and compared fully
developed circumbinary disks and their properties with observational data.
That work extended previous computations by \citet{1996ApJ...467L..77A}
and \citet{1997apro.conf..792R} by solving explicitly a time dependent
energy equation including the effects of viscous heating and
radiative cooling.  We investigated both the structure and dynamics of
the disk as well as the gas flow in the close vicinity of the binary
star.  To that purpose we utilized a newly developed method which
enables us to cover the whole spatial domain called the Dual-Grid
method.  For the first time we performed long-time integration of the
complete system covering several hundred orbital periods of the binary
and compared the properties of the evolved systems with observational
data such as spectral energy distributions in the infrared and optical
bands and accretion rates estimated from observed luminosities.

In this paper we extend the model by accounting for irradiation
effects which are important for passive accretion disks such as the
disks of DQ Tau and AK Sco, according to observational
data. Additionally, refined observational data is available from
\citet{2003A&A...409.1037A} for the AK Sco system which
allows to better constrain the parameters of the system by our
numerical simulations.

In the next section we present the equations that are being solved and the 
refined radiative balance model.
The overall layout of the physical model is the same as in
\citet{2002A&A...387..550G}. Then we proceed with some remarks on
the generation of the spectra from the simulation data (\Sect{sec:spectra}).
After that in \Sect{sec:setup} we describe the two-dimensional simulation setup and
proceed to the new results in \Sect{sec:results}. We summarize and conclude in
\Sect{sec:conclusion}.

\section{Equations}
The evolution of the disk is given by the two-dimensional
($r, \varphi$) evolutionary Navier-Stokes equations for the
density $\Sigma$,
the velocity field $\vec{u}\equiv (u_r, u_\varphi)$,
and the temperature $T$.
In a coordinate-free representation the equations read
\beq
\label{eq:sigma}
 \pdoverdt{\Sigma} + \nabla \cdot \left( \Sigma {\vec{u}} \right) = 0
\eeq
\beq
\label{eq:momentum}
 \pdoverdt{\Sigma{\vec{u}}} 
   + \nabla\cdot\left(\Sigma \vec{u} \cdot \vec{u}\right) = 
   -\nabla P - \Sigma\nabla\Phi + \nabla\cdot{\bf T}
\eeq
\beq
\label{eq:temperature}
 \pdoverdt{\Sigma\varepsilon} 
 + \nabla \cdot \left[ \left( \Sigma\varepsilon 
       + P \right) {\vec{u}} \right] = - \Sigma \vec{u} \cdot \Phi
  - \nabla\cdot \left(\vec{F} - \vec{u} \cdot {\bf T} \right)
\eeq
Here $\varepsilon = c_{\rm v} T + 1/2 u^2$ is the specific total energy,
$P$ is the vertically integrated (two-dimensional) pressure
$P = R \Sigma T/\mu$
with the midplane temperature $T$ and the mean molecular weight $\mu$,
which can be obtained by solving the Saha rate equations for Hydrogen dissociation
and ionization and Helium ionization. $\vec{F}$ and ${\bf T}$ are the
radiative flux and viscous stress tensor, respectively.

The gravitational potential $\Phi$ generated by the binary stars
is given by
\beq
 \label{eq:potential}
 \Phi = -\sum_{i=1,2}{\left\{\begin{array}{l@{\quad}l}
\frac{\displaystyle G\,M_i}{\displaystyle 
  |{\vec{r}} - {\vec{r}_i}|} & 
     {\rm for}\;\;|{\vec{r}} - {\vec{r}_i}|>R_* \\[1em]
\frac{\displaystyle G\,M_i\left(3R_*^2
    -|{\vec{r}} - {\vec{r}_i}|^2\right)}
     {\displaystyle 2 \, R_*^3} & 
   {\rm for}\;\;|{\vec{r}} - {\vec{r}_i}| \le R_*
\end{array}\right.}
\eeq
where $\vec{r}_i$
are the radius vectors to the two stars, and $R_*$ is the stellar radius
assumed to be identical for both stars.
The second case in Eq.~(\ref{eq:potential}) gives the potential inside
a star with radius $R_\star$ having a homogeneous (constant) density.
\subsection{Viscosity and Disk Height}
The effects of viscosity are contained in the viscous stress tensor
${\mathbf T}$. Here we assume that the accretion disk may be described
as a viscous medium driven by some internal turbulence which
we approximate with a Reynolds ansatz for the stress tensor.
The components of ${\mathbf T}$
in different coordinate systems are spelled out explicitly for
example in \citet{1978trs..book.....T}.
A useful form for disk calculations considering angular momentum
conservation is given in \citet{1999MNRAS.303..696K}.

For the kinematic shear viscosity $\nu$ we use
an $\alpha$-model \citep{1973A&A....24..337S} of the form
\beq
      \nu = \alpha c_{\rm s} H\,,
\eeq
where $c_{\rm s}$ and $H$ are the local sound speed and vertical height,
respectively. The local disk height $H(\vec{r})$ is computed from
the vertical hydrostatic equilibrium
which yields
\beq
  \label{eq:diskheight}
  H({\vec{r}}) = 
  \left(\sum_{i=1,2}{\frac{G M_i}
     {c_{\rm s}^2\left|{\vec{r}} - {\vec{r}_i}\right|^3}}
  \right)^{-\frac{1}{2}}
   =  \left(\sum_{i=1,2}{H_i^{-2} (\vec{r})}
  \right)^{-\frac{1}{2}}\,.
\eeq
This can be split into the single star disk heights given
by $H_i({\vec{r}})
=c_{\rm s}\sqrt{\frac{\left|{\vec{r}}-{\vec{r}_i}\right|^{3}}{G M_i}}$.
The sound speed is given here by $c_{\rm s} = R T / \mu$.
\subsection{Radiative balance}
The influence of radiative and viscous effects on the disk
temperature is treated in a local vertical balance.
The balance equation for the internal and radiative energy considering
these effects reads
\beq
  \label{eq:energybalance}
  \pdoverdt{\left(\Sigma c_{\rm V} T\right)} 
       = Q_{\rm diss} - Q_{\rm rad} + Q_{\rm stars}\,,
\eeq
where $Q_{\rm diss}$ and $Q_{\rm rad}$ denote the viscous dissipation
and radiative losses. $Q_{\rm stars}$ is the irradiation heating
by the stars, which is considered for the optically thick regions of
the circumbinary disk. This inclusion of irradiation effects
is done in a simple geometric fashion due to the lack of explicit treatment
of the vertical structure of the disk in these two-dimensional calculations.
For the viscous dissipation we use the vertically averaged expression
\beq
     Q_{\rm diss} = {\vec{u} \cdot \nabla {\mathbf T}} 
    = \frac {1}{2 \Sigma \nu} {\rm Tr}\left({\mathbf T}^2 \right)\,,
\eeq
where ${\mathbf T}$ is the viscous stress tensor.
For the radiative transport we use
\beq
        Q_{\rm rad} = - \nabla \cdot {\vec{F}_0} - \pdoverd{F_z}{z}
\eeq
where $\vec{F}_0$ is the flux vector in the $z=0$ plane.
Here we consider only the losses in the vertical direction,
i.e. $\vec{F}_0 =0$, a standard approximation in accretion disk theory.
Integration over the vertical direction yields
\beq
        Q_{\rm rad} = 2 F_{\rm rad} = 2 \sigma_{\rm B} T^4_{\rm eff}
\eeq
with the local effective (surface) temperature $T_{\rm eff}$.
Following \citet{1990ApJ...351..632H} this temperature is related
to the midplane temperature $T$ by the 
LTE gray solution for the vertical structure of accretion disks.
Simplifying this relation we get
\beq\label{eqn:hubeney}
       T^4 = \frac{3}{4} T_{\rm eff}^4 \left[\frac{1}{2}\tau + \frac{1}{\sqrt{3}} + \frac{1}{3\tau}\right]\,.
\eeq
The optical depth is calculated via \citep{1991ApJ...375..740R}
\beq
       \tau = \frac{1}{2}\kappa\Sigma
\eeq
using an interpolated opacity $\kappa(\rho, T)$ adapted from
\citet{1985prpl.conf..981L}. 
\subsubsection{Irradiation}
For the heating from the stars we use
\beq
       Q_{\rm stars} = A \sigma_{\rm B} T_*^4 \sum_i{\left(\frac{R_*}{r_i}\right)^2}\,,
\eeq
where $T_*$ is the effective blackbody temperature of the stars, $R_*$ is
the stellar radius and $r_i$ are the distances of the stars to the local patch.
The factor $A$ honors the effective surface the flux operates on. 
This is obtained by reconstructing a surface from the radially varying
pressure scale height.  Self-shadowing is
taken into account by setting $A$ to zero for back-faces.

Numerical tests have shown that
with the inclusion of irradiation effects we observe oscillatory
behavior of the temperature near to the edge of the gap due to self-shadowing
effects as also found by \citet{2000A&A...361L..17D}.  But we can
overcome this specific instability by using appropriate time sub-stepping for
the radiative balance. This suggests that if the radiative balance is
imposed using a time relaxation process the self-shadowing instability does not
occur. This may be either because the cooling timescale is not
within the region of linear instability or because of non-linear
effects exposed by the relaxation mechanism.

\section{Spectra Generation}
\label{sec:spectra}
Spectral energy distributions from the disk emission can be computed by integrating the flux
of the blackbody radiation from the disk surface. This is done decoupled
from the numerical simulation as a postprocessing step. We follow
\citet{1988ApJ...326..865A}, and compute the flux at the observer
via
\beq
  \label{eq:sed}
  F_\nu = \frac{\cos i}{D^2}\int_{r_0}^{R_d}\!\!\int_{0}^{2\pi}\!\!B_\nu\left[T(r,\varphi)\right]\left(1-e^{-\tau(r,\varphi)}\right)\,r{\rm d}\varphi\,{\rm d}r
\eeq
where $i$ is the inclination of the disk, $D$ the distance to the observer
and $\tau$ the line-of-sight optical depth through the disk, which can
be approximated using the opacity $\kappa$, $\tau(r,\varphi)=\frac{\kappa \Sigma(r,\varphi)}{\cos i}$. The surface disk temperature $T$ is
estimated using the optical depth of the disk as in the
radiative balance process.

This integration is limited to regions outside of the stellar cores as
temperatures and densities inside the cores are not modeled correctly.
The stellar emission is accounted for by adding two blackbody spectra
for appropriate effective stellar temperatures. To compare with observational
data all of the generated spectra are reddened
according to an extinction $A_v$ suggested by the cited references using the method
from \citet{1989ApJ...345..245C}.

\section{Simulation Setup}
\label{sec:setup}
\begin{figure}
\begin{center}
\resizebox{0.9\linewidth}{!}{%
\includegraphics{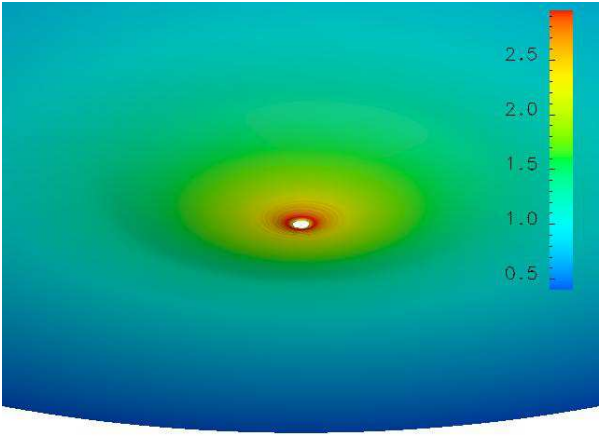}}
\end{center}
\caption{Visible setup of the AK Sco disk system using the 
estimated inclination of $63^\circ$.
The disk shape is modeled according
to pressure scale height, gray scale coding is logarithmic surface density in ${\rm g}/{\rm cm}^2$. The disk extent is $12\,$AU.}
\label{fig:akscoincl}
\end{figure}

We start with an axisymmetric circumbinary disk around
the binary system with an initial gap of a width that is determined by simulations
including the gap formation process. This results in gap widths about three to four times the binary separation
radius consistent with results from \citet{1994ApJ...421..651A}. This initial configuration is then evolved for several orbital periods, so the
circumstellar environment can form and the whole system settles into
a quasi-periodic state.

Initial density and temperature distributions for the disk are taken from power-law
fits of \citet{1997AJ....113.1841M} and
\citet{2003A&A...409.1037A}. This initial surface temperature is relaxed to a
midplane temperature satisfying Eq.~(\ref{eqn:hubeney}).

The stellar temperatures and radii specify the flux observed, and
that transfered into the disk.
These parameters cannot be fitted independently but
need to be fixed for the irradiation of the disk.
This is why we use stellar radii as suggested by
\citet{1997AJ....113.1841M} and \citet{2003A&A...409.1037A} and fit
the effective temperatures $T_*$ according to the observed spectral energy
distributions by assuming simple blackbody spectra. 
Direct  matching of the stellar spectra is a
good approximation as can be seen in Fig.~\ref{fig:akscoincl}, where we
display the AK Sco disk with the binary stars projected using the
estimated inclination of 63$^\circ$. This demonstrates that occultation of the
stars by the circumbinary disk is not to be expected.

\begin{figure}
\begin{center}
\resizebox{0.9\linewidth}{!}{%
\includegraphics{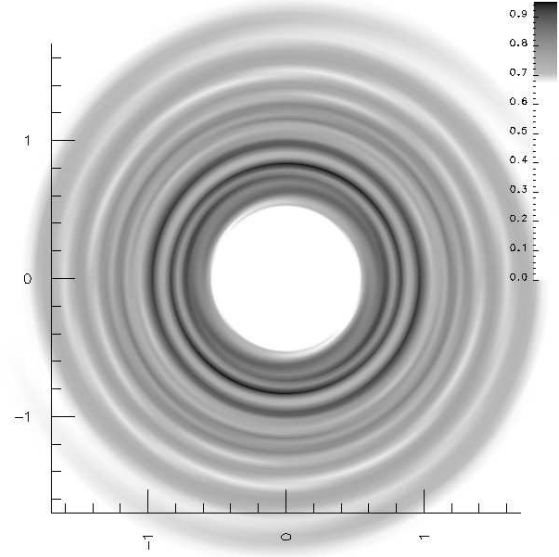}}
\end{center}
\caption{Linear surface density (${\rm g}/{\rm cm}^2$) of the inner
circumbinary disk of the DQ Tau system after $50$ orbital periods.
Scaling is in AU.
The spiral structure in the circumbinary disk induced by the
binary is clearly seen.}
\label{fig:dqtaucbspiral}
\end{figure}

The overall azimuthally averaged circumbinary disk shape does not change
significantly on timescales of the orbital
period of the binary system. Induced by the gravitational torques
of the central binary a nice a spiral structure develops in the
inner regions of the disk as can be seen in Fig.~\ref{fig:dqtaucbspiral}.

\begin{figure}
\begin{center}
\resizebox{0.9\linewidth}{!}{%
\includegraphics{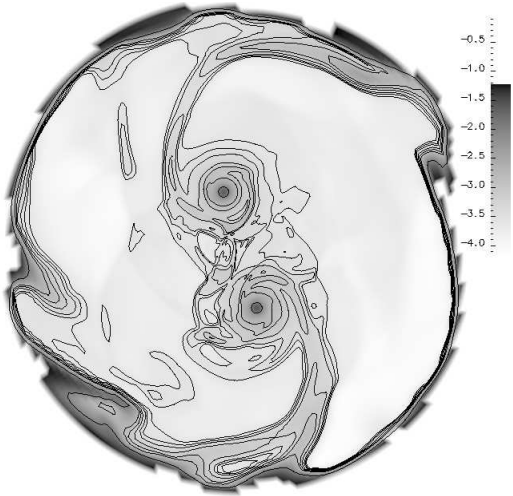}}
\end{center}
\caption{The spiral arms onto the circumstellar environment for the
DQ Tau system after $50$ orbital periods.
The gray scale follows logarithmic surface density annotated with iso-lines.
Plotted is the region inside the circumbinary disk gap which is located at $0.4\,$AU. The binary
is in apastron phase.}
\label{fig:dqtauspiral}
\end{figure}

Also one can observe periodic formation of
spiral arms from the edge of the circumbinary disk going down to the
circumstellar environment (Fig.~\ref{fig:dqtauspiral}) connected to the
binary orbital period.
The quasi-periodic state of the {\it circumstellar} environment of the
eccentric system is characterized by the following phases:
\begin{enumerate}
\item Going from periastron to apastron, material is accreted to
  the circumstellar environment from the circumbinary disk and the remaining
  material in the gap.
\item Around apastron phase circumstellar envelopes/disks have been formed.
\item Approaching periastron, these circumstellar disks are torn off due to
  gravitational torques exerted on the disks by the stars, and a part of the
  mass is accreted onto the stars.
\end{enumerate}
This leads to periodic accretion events which is both confirmed from observations
\citep{1997AJ....114..781B} showing periodic brightening events and from numerical simulations.
\noindent A more detailed discussion on the disk dynamics and accretion can be found in
\citet{2002A&A...387..550G}.

\subsection{Dual-Grid and SPH}
\label{subsec:SPH}
We additionally have performed Smoothed Particle Hydrodynamics (SPH) test calculations
which cover only the inner part of the simulation domain
to corroborate the results of the Dual-Grid technique.
The initial setup consists of two circumstellar disks in
Keplerian motion around the individual stars in apastron.  
Similar to the grid
based calculations we evolve this system for a couple of orbital periods to
reach a quasi-stationary state.  For detailed
descriptions of the SPH method see, e.g., \cite{1990nmns.work..269B} and 
\cite{Monaghan:1992:Sph}.  The SPH simulations include only radiative
cooling and viscous heating and not the irradiation from the stars, as
these calculations are primarily intended to support the results obtained
by the Dual-Grid technique.

\section{Results}
\label{sec:results}

Grid based calculations have been performed on two different regions of the
whole system. First, high resolution ($110\times 204$) calculations extending only
over the circumbinary disk ($0.4$--$12\,$AU for AK Sco and $0.3$--$80\,$AU for DQ Tau)
have been performed to evolve these regions in higher resolution and for a long
time. Second, high resolution 
($180\times 247$ for the $r-\varphi$ grid, $101\times 101$ for the Cartesian grid)
calculations extending over the circumstellar disks and up to the very inner
region of the circumbinary disk (up to $1.2\,$AU for AK Sco and up to $1.0\,$AU for DQ Tau) 
have been run to allow comparisons with the SPH
calculations.
Finally, low resolution ($128\times 91$, $37\times 37$) simulations of the
whole system using the Dual-Grid technique have been performed to
show that indeed these regions decouple in the timescales we are interested in.
\subsection{Dual-Grid and SPH}
\label{sec:dualsph}
The Dual-Grid technique was developed to overcome limitations of using
cylindrical coordinates for accretion disks, where the whole domain
including the origin needs to be covered
\citep{2002A&A...387..550G}. To demonstrate that the interpolation scheme
used between the Cartesian and the cylindrical grid is
accurate enough for modeling close binary circumstellar disks, we performed
comparison runs using SPH.

Simulations are set up to initially contain two circumstellar disks around
the individual stars with a mass of $10^{-8}\,\MSOL$ and a disk profile
according to $\Sigma \propto r^{-3/2}$. This system is then evolved
for a few orbital periods to truncate the disks to the right size and go into
a quasi-stationary equilibrium.
Dual-Grid calculations were performed on a $180\times 247$ sized $r-\varphi$ grid and a $101\times 101$ sized Cartesian grid, while the SPH calculations instrumented $150\,000$ particles.

\begin{figure}
\resizebox{1.0\linewidth}{!}{%
\includegraphics{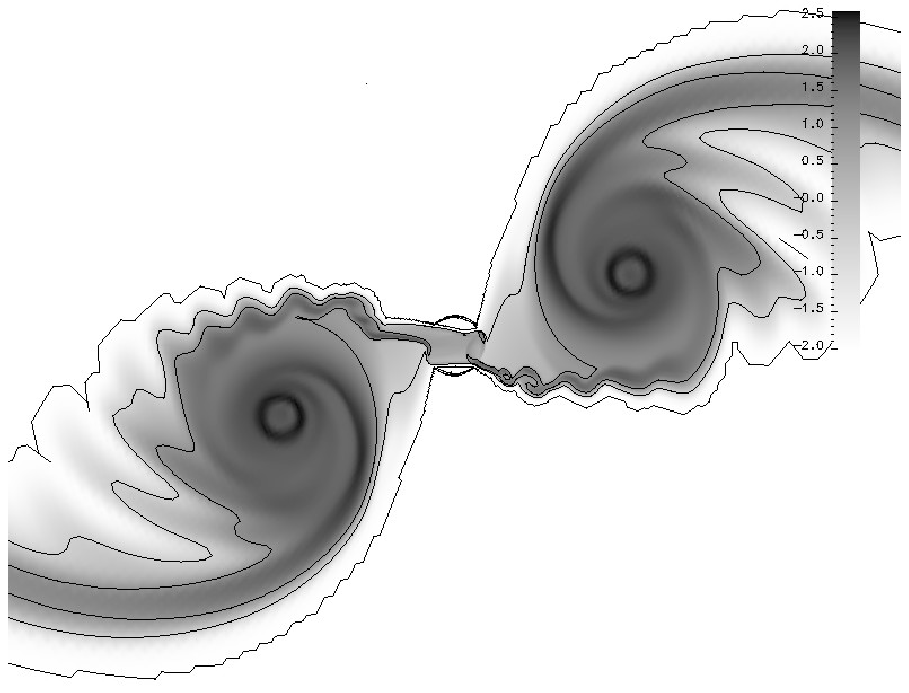}%
\includegraphics{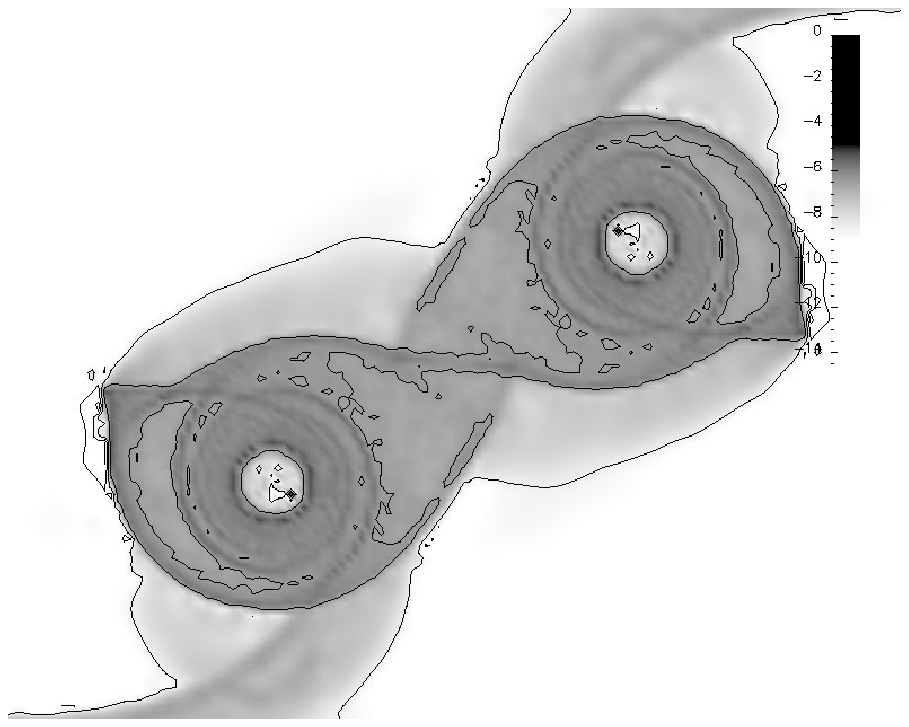}}
\caption{Circumstellar disks logarithmic density structure for Dual-Grid simulation (left)
and SPH simulation (right) while approaching periastron.  The binary separation is about
$0.1\,$AU.}
\label{fig:cscirc}
\end{figure}

Fig.~\ref{fig:cscirc} shows the structure of the disks for Dual-Grid and
SPH calculations after the same evolution time. In both simulations one can see similar
circumstellar material features, namely
\begin{itemize}
\item spiral structure inside the circumstellar material,
\item spiral waves going off the circumstellar region,
\item a bar-like connection between both circumstellar regions.
\end{itemize}
\noindent Thus we can conclude that with the Dual-Grid calculations we see the same features
across the origin as with the SPH calculations. The Dual-Grid technique is applicable to simulation of
circumbinary systems including the origin and the inner circumstellar regions.

\subsection{Irradiation Effects}
\label{sec:irradiation}

\begin{figure}
\begin{center}
\resizebox{1.0\linewidth}{!}{%
\includegraphics{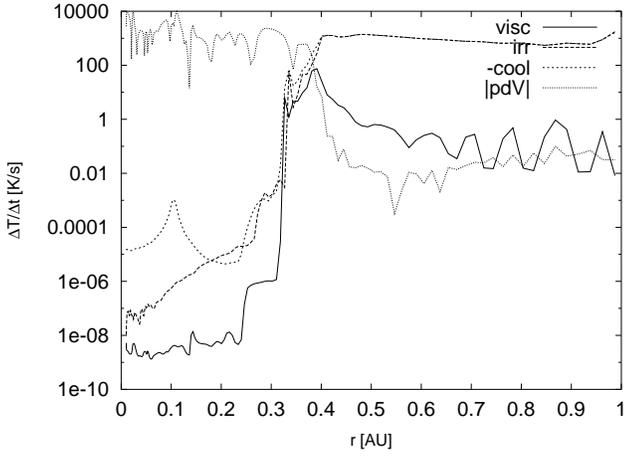}}
\end{center}
\caption{
Contribution of the different heating/cooling terms
to the rate of temperature change 
(DQ Tau after 50 orbital periods).
Only the inner part of the system
is displayed. The irradiation {\sf irr}
and the radiative cooling {\sf -cool} match exactly beyond
$0.4\,$AU.}
\label{fig:dqtauradi}
\end{figure}

\begin{figure}
\begin{center}
\resizebox{1.0\linewidth}{!}{%
\includegraphics{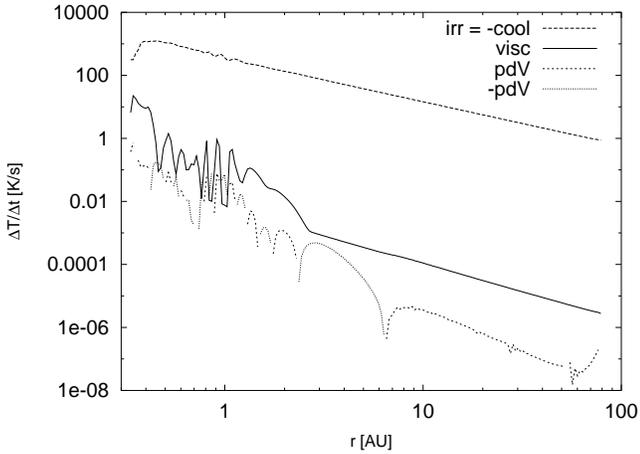}}
\end{center}
\caption{
Contribution of the different heating/cooling terms
to the rate of temperature change 
(DQ Tau after 50 orbital periods).
Plotted is the outer circumbinary disk part of the system. 
The irradiation by the central stars and
the radiative cooling match exactly (topmost dashed line).}
\label{fig:dqtaurado}
\end{figure}

For motivating the inclusion of irradiation effects,
we first analyze the individual contribution of the different terms in the radiative
balance (Eq.~\ref{eq:energybalance}), and the pressure work $p \nabla \cdot {\bf u}$,
to the temperature change rate.
The azimuthally averaged rate of the temperature change at quasi-stationary equilibrium of
the DQ Tau system is plotted in Fig.~\ref{fig:dqtauradi} for
the inner regions of the disk and Fig.~\ref{fig:dqtaurado} for the
outer regions. Here, {\sf visc} denotes the rate from the viscous heating,
{\sf irr} the rate from the irradiation process, {\sf cool} the emission contribution
and {\sf pdV} the rate from the pressure work.
Note, that the inner edge of the disk is approximately
at $r = 0.4\,$AU for this model.

As one can easily infer from Fig.~\ref{fig:dqtaurado} the circumbinary
disk is dominated by irradiation effects (which in fact match the
cooling rates exactly), and as such qualifies as a passive disk.
The effects due to viscous heating and pressure work can be neglected
here.

In the inner optically thin part the pressure work is dominating due
to the dynamic motion in this region caused by the gravitational
torques exerted by the eccentric binary. One can also see that the
emitted flux in this region is negligible
compared to the flux emitted by the circumbinary disk.

\subsection{Circumbinary Disks}
\label{sec:cbdisk}

We model the two close spectroscopic binary systems with circumbinary
disks, DQ Tau and AK Sco, which are both of T Tauri type. The physical
parameters for the systems are taken from
\citet{2003A&A...409.1037A} for AK Sco
and \citet{1997AJ....113.1841M} for DQ Tau.
The relevant information for the simulations has been
summarized in Tables~\ref{table:spectroscopic} and \ref{table:spectroscopicak}.
Both stars are assumed to have the same effective temperature $T_*$ and the
same radius $R_*$.

For both the DQ Tau and the AK Sco system we shall
see missing flux at around $5\times 10^{13}$, resp.~$10^{14}\,$Hz which can
be accounted to the modeling of the T Tauri type stars emission as a
simple blackbody.

\begin{table}
\begin{center}
\begin{tabular}{ll}\hline\hline
$P$            & $15^d.8043\pm0^d.0024$ \\
$e$            & $0.556\pm0.018$ \\
$i$            & $23^\circ$ \\
$a\sin i$      & $0.024\pm0.009\,\AU$ \\
$M_{\rm A}\sin^3i$          & $0.033\pm0.002\,\MSOL$              \\
$M_{\rm B}\sin^3i$          & $0.033\pm0.002\,\MSOL$              \\
$q$            & $1.0\pm0.03$ \\
$T_*$          & 4000\,K \\
$R_*$          & $1.785\,\RSOL$ \\
$d$            & $140\,\PC$ \\
$M_{\rm d}$    & $0.002-0.02\,\MSOL$ \\
$R_{\rm d}$    & $50\,\AU$ \\
$A_{v}$        & $2.0$ \\ \hline
\end{tabular}
\end{center}
\caption{Parameters for \object{DQ Tau} \citep{1997AJ....113.1841M}.}
\label{table:spectroscopic}
\end{table}

\begin{table}
\begin{center}
\begin{tabular}{ll}\hline\hline
$P$            & $13^d.609453\pm0^d.000026$ \\
$e$            & $0.4712\pm0.0020$          \\
$i$            & $63^\circ$                \\
$a\sin i$      & $0.14318\pm0.00005\,\AU$               \\
$M_{\rm A}\sin^3i$          & $1.064\pm0.007\,\MSOL$              \\
$M_{\rm B}\sin^3i$          & $1.050\pm0.007\,\MSOL$              \\
$q=M_{\rm B}/M_{\rm A}$            & $0.987\pm0.005$                  \\
$T_*$          & $6500\pm100$\,K                   \\
$R_*$          & $1.59\pm0.35\,\RSOL$              \\
$d$            & $152\,\PC$                \\
$M_{\rm d}$    & $0.005\,\MSOL$            \\
$R_{\rm d}$    & $40\,\AU$                 \\
$A_{v}$        & $0.5\pm0.1$                      \\ \hline
\end{tabular}
\end{center}
\caption{Parameters for \object{AK Sco} \citep{2003A&A...409.1037A}.}
\label{table:spectroscopicak}
\end{table}

\subsubsection{DQ Tau}
\label{subsec:dqtau}

\begin{figure}
\begin{center}
\resizebox{1.0\linewidth}{!}{%
\includegraphics{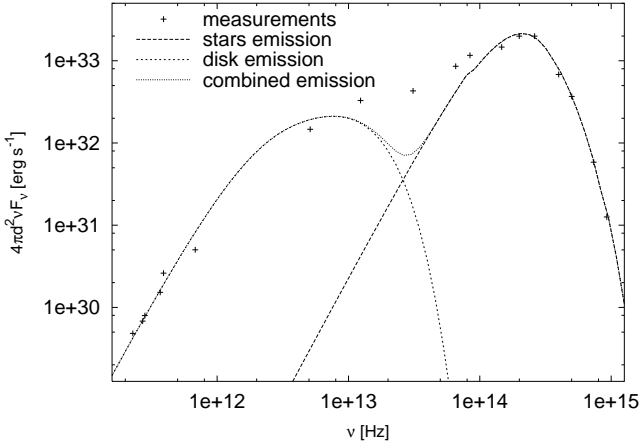}}
\end{center}
\caption{Computed SED of DQ Tau.
Crosses show observational data taken from \citet{1997AJ....113.1841M}.}
\label{fig:dqtaused}
\end{figure}

\begin{figure}
\begin{center}
\resizebox{1.0\linewidth}{!}{%
\includegraphics{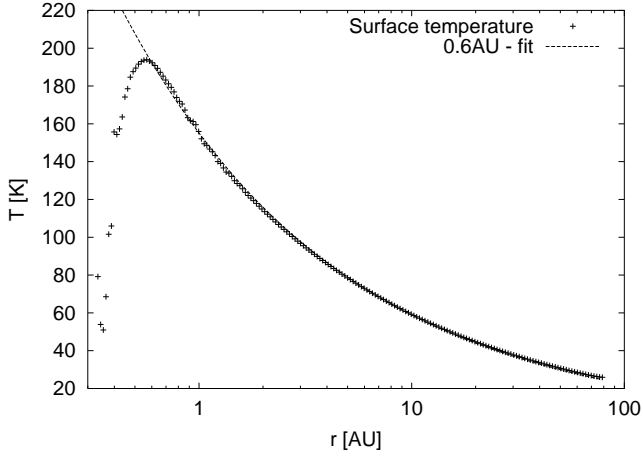}}
\end{center}
\caption{Azimuthally averaged surface temperature of the equilibrium
DQ Tau circumbinary disk.}
\label{fig:dqtautemp}
\end{figure}

From full simulations including the gap formation process we infer an initial gap with
a radius of $0.4\,\AU$ for the DQ Tau system.
The stars combined effective blackbody temperature is fitted manually to $3500\,{\rm K}$
to match the observations.
Starting from the parameters presented in Table~\ref{table:spectroscopic} and power-law
fits from \citet{1997AJ....113.1841M} we adjust
the initial density distribution power-law index, the disk mass and its extension to best
match the observed SED after going to quasi-stationary state.

The SED of DQ Tau can be fitted best with a thin disk extending
to $80\,$AU which has an unusually
flat surface density distribution $\Sigma \propto r^{-0.25}$
and a total disk mass of $7\times 10^{-4}\,\MSOL$. This
is less mass and a larger disk radius as obtained by \citet{1997AJ....113.1841M}.
After fifty orbital periods,
the equilibrium surface temperature distribution of the circumbinary
disk follows $T(r) = 155\,{\rm K}\;r^{-0.42}$ as can be seen in Fig.~\ref{fig:dqtautemp}.

\subsubsection{AK Sco}
\label{subsec:aksco}

\begin{figure}
\begin{center}
\resizebox{1.0\linewidth}{!}{%
\includegraphics{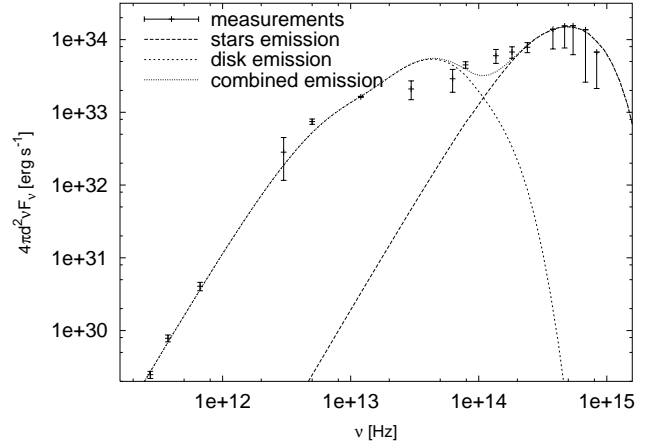}}
\end{center}
\caption{Computed SED of AK Sco with parameters and observational data
(crosses with error-bars) taken from \citet{2003A&A...409.1037A}.}
\label{fig:akscoseda}
\end{figure}

In Fig.~16 of \citet{2003A&A...409.1037A} the authors propose a circumbinary disk mass of
$M_{\rm d} = 0.02\,\MSOL$ and a disk extend of $R_{\rm d} = 12\,\AU$ for the best
fit to the observed SED. The surface density distribution is that of a minimum mass solar nebula
as in the used model \citep{1997ApJ...490..368C}, so $\Sigma(r) \sim r^{-1.5}$.
In Fig.~\ref{fig:akscoseda} the emission of a model evolved with these parameters
imposed initially is displayed.

\begin{figure}
\begin{center}
\resizebox{1.0\linewidth}{!}{%
\includegraphics{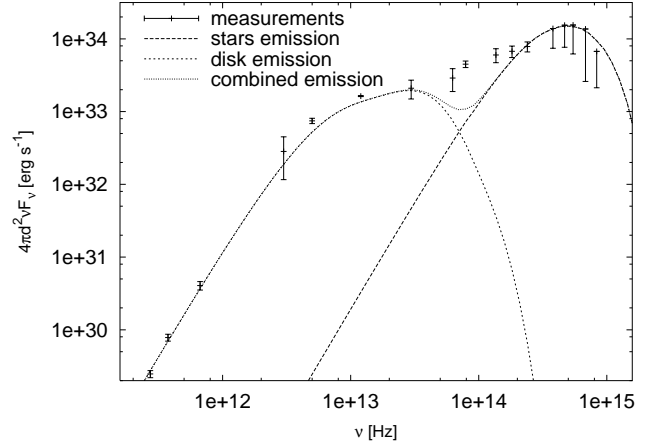}}
\end{center}
\caption{Computed SED of the AK Sco system with best-fit parameters.
Crosses with error-bars show observational data taken from \citet{2003A&A...409.1037A}.}
\label{fig:akscosed}
\end{figure}

Clearly visible is the excess flux of the model in the infrared (around $5\times 10^{13}\,$Hz)
which we can reduce by lowering the
disk mass and slightly flattening the disk density profile. Using a disk mass of
$M_{\rm d} = 0.01\,\MSOL$ and an initial density profile according to $\Sigma(r) \sim r^{-1.0}$
we obtain a spectrum as seen in Fig.~\ref{fig:akscosed}.

\begin{figure}
\begin{center}
\resizebox{1.0\linewidth}{!}{%
\includegraphics{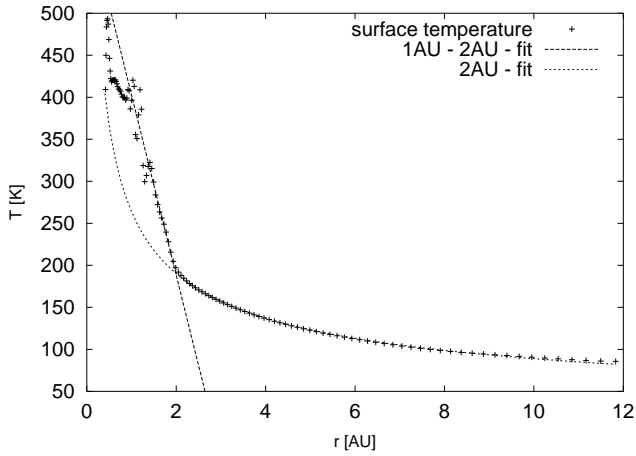}}
\end{center}
\caption{Azimuthally averaged surface temperature of the
equilibrium AK Sco circumbinary disk.}
\label{fig:akscotemp}
\end{figure}

For the AK Sco system the stars combined effective blackbody temperature is best fitted
to the observations by assuming $T_* = 6500\,$K. In a quasi-stationary
state after about fifty orbital periods of the binary the surface temperature can be fitted
to the power-law $T(r) = 264\,{\rm K}\;r^{-0.47}$ for $r>2\,$AU and a surprisingly steep linear
behavior in the region between one and two AU ($T(r) = 617\,{\rm K}-215\,{\rm K}\;r$)
as shown in Fig.~\ref{fig:akscotemp}.  This is due to the different opacities in the dense
region of the disk and compared to the case of DQ Tau three orders of magnitude higher
maximum surface density which is related to the AK Sco disk being much smaller and more massive.

We note that our models are not able to fit the higher flux observed in the SED
of both systems in the region between the disk and stellar contributions.
This may be attributed to line emission features \citep{2003A&A...409.1037A}
which cannot be modeled by our simulations.

\subsection{Circumstellar Disks}
\label{sec:csdisk}

For the region inside the circumbinary disk gap we see a tiny amount
of warm gas and streamers feeding circumstellar disk like structures
from the inner circumbinary disk edge. Due to the low mass this material
does not contribute to the continuum part of the observed spectral
energy distributions but rather would show up in lines and the UV part
of the spectra.

Also, the observed periodic brightening in the light-curve from the
systems originate here, which is believed to originate from periodic
accretion events. These periodic events of accretion have been confirmed
by \citet{1996ApJ...467L..77A}, \citet{1997apro.conf..792R} and
\citet{2002A&A...387..550G} through numerical simulations.

\begin{figure}
\begin{center}
\resizebox{1.0\linewidth}{!}{%
\includegraphics{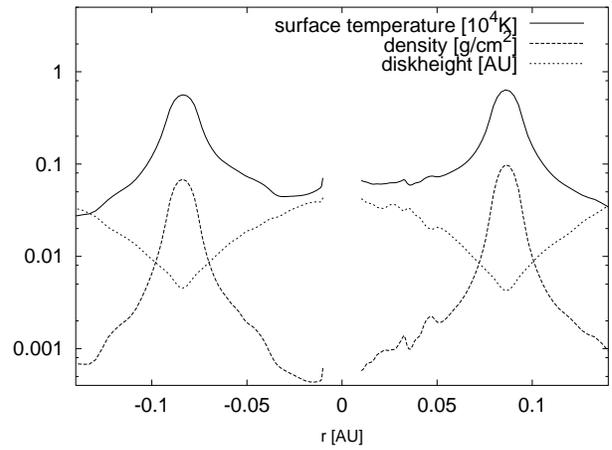}}
\end{center}
\caption{Profiles of the circumstellar material along the connecting
line of the binary at apastron in the system of DQ Tau (stellar radii $0.008\,\AU$).}
\label{fig:csdiska}
\end{figure}

\begin{figure}
\begin{center}
\resizebox{1.0\linewidth}{!}{%
\includegraphics{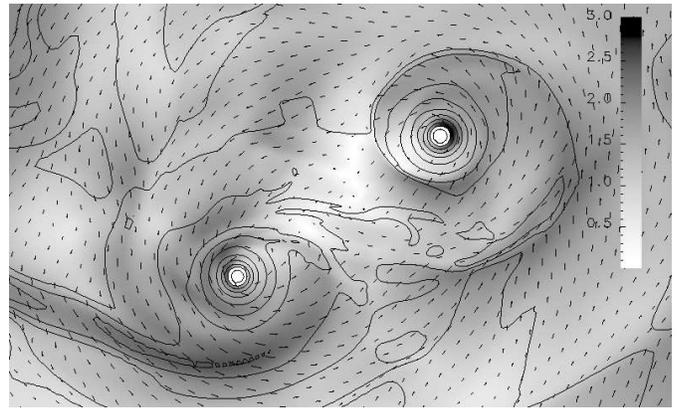}}
\end{center}
\caption{Circumstellar material in the DQ Tau system after 41 orbital periods,
corresponding to Fig.~\ref{fig:csdiska}.
Gray scale coding is velocity magnitude annotated with vector glyphs,
isolines are equally spaced logarithmic surface density. The stellar cores are the white
circles.}
\label{fig:csdiskc}
\end{figure}

The circumstellar disk like structures around DQ Tau form through
accumulation of material in the stellar gravitational potential. Without
removing any material through an accretion process, disk profiles according
to Fig.~\ref{fig:csdiska} form.
They cannot be identified with a
classical accretion disk due to shape and size as one can see from
Fig.~\ref{fig:csdiskc} which shows the circumstellar material and velocity distribution.
They rather would form sort of an envelope,
but this remains to be investigated in three-dimensional calculations.

\section{Conclusion}
\label{sec:conclusion}
We model the
circumbinary disks of DQ Tau and AK Sco
and their circumstellar environment including emitted spectra by
applying two-dimensional
hydrodynamical simulations of high accuracy and resolution.
The models include a self-consistent approximative vertical energy transport
including irradiation from the stars
using detailed opacities and equation of state.
The numerical resolution we have reached by now is so fine that the
individual stars in close binaries are already resolved by typically about
$10\times 10$ grid-cells for high-resolution calculations.

We find that with suitable variations of
\begin{enumerate}
\item  the initial surface density profile $\Sigma \sim r^{-d}$,
\item  the circumbinary disk mass $M_{\rm d}$
\item  and the circumbinary disk extent $R_{\rm d}$
\end{enumerate}
the spectral energy distributions of DQ Tau and AK Sco
can be fitted well to the observational data only if
irradiation effects are included.

The circumbinary disk settles into a quasi-stationary state after around 50 orbital
periods of the binary with a temperature distribution following approximately
$T(r) \propto r^{-0.45}$ for larger radii in both systems. 
We find that irradiation plays the dominant role for the heating balance
in these outer regions. In the inner part of the disk just beyond the
inner gap, the additional heating due to the tidal effects is important,
particularly for small disks around tight binaries, such as AK Sco. 
 
For the present model parameter we do not find any indication for
a previously observed self-shadowing instability of the disk due to
the stellar irradiation \citep{2000A&A...361L..17D}.
A proper numerical treatment of the strongly non-linear cooling
is required to prevent artificial numerical instabilities, which is achieved
by using a time-relaxation scheme as opposed to directly solving for the
equilibrium.

Future models of this kind need to include a more detailed treatment of the stellar
boundary layers as well as a more detailed model of the accretion process.
We believe that doing full radiative transfer in two dimensions will not
improve our results, but going to three dimensions with full radiative
transfer included will possibly enable us to obtain more accurate spectra from the
circumbinary disk gap region with its thin material.
We then may be able to compare in more detail
with observed light curves and phase dependent spectra.
\begin{acknowledgements}
This work was funded by the German Science Foundation (DFG) under SFB 382
{\it Simulation physikalischer Prozesse auf H\"ochstleistungsrechnern}.
\end{acknowledgements}
%
\bibliographystyle{aa}
\bibliography{0223}
\end{document}